\documentclass[aip,
 amsmath,amssymb,
 reprint,%
groupedaddress, 
]{revtex4-2}

\usepackage{graphicx}
\usepackage{dcolumn}
\usepackage{bm}

\usepackage[utf8]{inputenc}
\usepackage[T1]{fontenc}
\usepackage{mathptmx}
\usepackage{amsmath}
\usepackage{xcolor}
\usepackage{soul}

\usepackage{graphicx}
\usepackage{amsmath}
\usepackage{subcaption}
\usepackage{multirow}
\usepackage{ragged2e}
\usepackage{booktabs,caption}
\usepackage{lineno,hyperref}
\usepackage[flushleft]{threeparttable}
\usepackage{adjustbox}
 \usepackage[final]{changes}

\begin{document}

\preprint{AIP/123-QED}

\title{Measurement of Third-Order Elastic Constants Using Thermal Modulation of Ultrasonic Waves}

\author{Bibo Zhong}
\email{bbzhong@huskers.unl.edu}
\author{Jinying Zhu}%
 \email[Author to whom correspondence should be addressed:]{jyzhu@unl.edu.}
\affiliation{ 
Department of Civil and Environmental Engineering, University of Nebraska-Lincoln, 1110 S 67th St. Omaha, NE 68182, USA
}%

\date{\today}

\begin{abstract}
Third-order elastic constants (TOEC) play an important role in nonlinear material characterization, but measurements of TOEC are laborious with large error margins. \replaced{This letter presents the equations of wave velocity changes caused by homogeneous temperature variation and uniaxial stress in isotropic media, and the expression of TOEC in terms of thermally induced velocity change and thermal strain}{This letter presents theoretical derivations of wave velocity changes caused by homogeneous temperature variation and uniaxial stress in isotropic media, and derives the TOEC using thermally induced velocity change and thermal strain}. TOEC of an aluminum sample were experimentally determined by measuring ultrasonic wave velocity changes in the uniaxial loading test and thermal modulation test. Experimental results showed good agreement between the two test methods. Owing to the simple test setup and high measurement sensitivity, the thermal modulation test is a potential experimental method to determine TOEC and absolute acoustic nonlinearity parameters.
\end{abstract}

\maketitle


The third-order elastic constants (TOEC) play an important role in nonlinear material characterization. In engineering applications, TOEC have been used to evaluate applied or residual stress based on the theory of acoustoelasticity \cite{hughes1953second, pao1985acoustoelastic,gandhi2012acoustoelastic}. Another application exploits the high sensitivity of TOEC in the evolution of different microstructural features, such as fatigue and dislocation related damage and thermal aging damage \cite{matlack2015review}.

In recent years nonlinear acoustic techniques have attracted considerable attention in nondestructive evaluation (NDE) owing to their high sensitivity to microscopic defects that often remain hidden to linear techniques \cite{matlack2015review,kundu2018nonlinear}.
In particular, \replaced{for a one-dimensional longitudinal wave (P wave) propagating in an isotropic material}{for a longitudinal wave (P wave) propagating in an one-dimensional medium}, the nonlinearity parameter $\beta$ can be expressed as \replaced{$\beta=-3-(2l+4m)/(\lambda+2\mu)$}{$\beta=3+(2l+4m)/(\lambda+2\mu)$} \cite{matlack2015review}. 
In most studies, nonlinear parameters are typically measured as relative values, since measuring the absolute values requires tedious calibration of the testing system and transducers and measures the absolute displacement amplitudes of ultrasonic waves and strains \cite{dace1991nonlinear}.

TOEC are typically determined based on the acoustoelastic theory \cite{hughes1953second,egle1976measurement} by measuring acoustoelastic coefficients, the change of elastic wave velocity under applied stress (or strain), through either uniaxial \cite{egle1976measurement} or hydrostatic \cite{daniels1958pressure} compression tests.  However, there are many challenges in these measurements. First, complicated equipment is required to produce hydrostatic pressure. Second, although uniaxial compression can produce sufficiently large strain and measurable wave speed change, it may change the dislocation network \cite{thomas1968third}, cause slip and plastic strain \cite{bateman1961third} in the materials, and lead to permanent change material properties.

In this letter, the authors present an analytical and experimental study to use thermal strain as a driving force for wave velocity change and measure the TOEC based on their relationships. This method was initially proposed by Sun and Zhu \cite{sun2020determination} to measure the absolute nonlinear parameters in metals and concrete and evaluate microcracking damage in concrete \cite{sun2019thermal}. They named this method as thermal modulation of nonlinear ultrasonic wave test. 

Compared to the uniaxial or hydrostatic tests, the experimental setup of the thermal modulation test is much simpler and can easily produce large and uniform strains in elastic media. In addition, as shown in this letter, thermally induced wave velocity changes are larger than mechanical strain induced velocity changes so that the velocity changes can be measured with less error. This method is naturally immune from thermal effect, since temperature variation is no longer an undesired influencing factor in tests. 
In the prior study \cite{sun2020determination}, the nonlinear parameter was derived based on a one-dimensional constitutive relationship. In this letter, TOEC are derived based on a three-dimensional elastic model, and experimental results from the thermal modulation test were compared and validated with the uniaxial loading test.


Acoustoelaticity describes the stress (strain) dependence of elastic wave velocity in a solid medium. Based on Murnaghan's finite deformation theory \cite{murnaghan1951finite}, Hughes and Kelly \cite{hughes1953second} derived the wave velocities under uniaxial and hydrostatic stresses in isotropic solids.  Pao and Gamer later \cite{pao1985acoustoelastic} re-examined the acoustoelastic theory and extended it to an infinitesimal strain superimposed on a predeformed body. Following their notations,  a body in a stress-free and strain-free state is called in the natural state while the body undergoing a static deformation is called the initial state. Coordinates of a material point in the natural states are represented by the position vectors \textbf{\textit{a}} (natural frame of reference). 
Pao and Gamer derived the equations of acoustoelaticity in both natural and initial frames of reference.

Although the solution of actual (initial) wave velocities are presented in most papers and widely cited \cite{hughes1953second,egle1976measurement}, there are many advantages to use the natural velocity in experiments, as discussed by Thurston and Brugger \cite{thurston1964third}. The natural velocity is calculated from the original travel length $L_0$ and the actual travel time $t$ as $V=L_0/t$.  Because $L_0$ does not change in the natural coordinates, the natural velocity is inversely proportional to the wave travel time, and then the relative velocity change can be calculated by measuring the relative time change, i.e., $dV/V^0=-dt/t^0$, where $V^0$ and $t^0$ are the velocity and travel time before deformation. While in the initial coordinates (deformed state), the actual velocity calculation needs correction for the actual travel length, which should be measured at each initial state \cite{pao1985acoustoelastic}. In experiments, the actual velocity is calculated as $dv/v^0=dL/L^0-dt/t^0$. In this letter, all equations and calculations are presented in natural coordinates, and natural wave velocities were used in derivations and experiments.

In the case of homogeneous predeformation, the equation of infinitesimal wave motion in an acoustoelastic medium is written as \cite{thurston1964third,pao1985acoustoelastic}:
\begin{equation}
\label{eq:equation of motion in natural coordinates}
    A_{ijkl} \frac{\partial^{2} u_{k}}{\partial a_{j} \partial a_{l}}=\rho^{0} \frac{\partial^{2} u_{i}}{\partial t^{2}}
\end{equation}
where $\rho^{0}$ is the density in the natural state (undeformed configuration), {\textit{u}} is the the dynamic displacement, or wave motion in the medium, $a$ is the position vector in the natural coordinates, and $t$ is the time. The elastic tensor $A_{ijkl}$ depends on second- and \replaced{third- order}{third-order} elastic constants, strains, and initial stresses.

Based on Pao and Gamer's derivations \cite{pao1985acoustoelastic}, Dodson and Inman \cite{dodson2014investigating} extended the acoustoelastic equations of motion to include thermal effects.  When an isotropic elastic body is subject to uniform temperature variation and has free boundary conditions, the initial strain is only caused by thermal expansion. Then the thermo-acoustoelastic tensor $A_{ijkl}$ can be written as \cite{dodson2014investigating}
\begin{equation}
\label{eq:A_T_simple}
    \begin{aligned}
        A_{i j k l}=&(\lambda+\alpha_{T} \Delta T(6 l-2 m+n+2 \lambda)) \delta_{i j} \delta_{k l} \\
        &+(\mu+\alpha_{T} \Delta T(3 m-\frac{1}{2} n+2 \mu))(\delta_{i k} \delta_{j l}+\delta_{i l} \delta_{j k}),
    \end{aligned}
\end{equation}
where $\alpha_T$ is the coefficient of thermal expansion, $\Delta T$ is the temperature change, $\delta_{ij}$ is the Kronecker delta function, the parameters $\lambda$ and $\mu$ are the second-order elastic constants known as Lam\'e constants, and $l$, $m$, $n$ are TOEC or Murnaghan constants.

Velocity changes of P and S waves induced by the thermal strain $\varepsilon_{T}=\alpha_T\Delta T$ can be derived from Eq. \eqref{eq:A_T_simple} as\cite{pao1985acoustoelastic}:
\begin{subequations}
\label{eq:Vp and Vs thermal expansion}
\begin{align}
    \rho_{0} V_{p}^{2}&= A_{1111}=\lambda+2 \mu+\epsilon_{T}(6 l+4 m+2\lambda+4\mu)\\
    \rho_{0} V_{s}^{2}&= A_{2121}=\mu+\epsilon_{T}(3 m-\frac{n}{2}+2\mu)
\end{align}
\end{subequations}
The thermo-acoustoelastic coefficients, defined as the relative changes of $V_p$ and $V_s$ by the thermal strain $\epsilon_{T}$, are related to TOEC by:
\begin{subequations}
\label{eq:dVp and dVs thermal expansion using l m n}
\begin{align}
    \frac{dV_p/V^{0}_p}{d\epsilon_{T}}&=1+\frac{3l+2m}{\lambda+2\mu}\\
    \frac{dV_s/V^{0}_s}{d\epsilon_{T}}&=1+\frac{6m-n}{4\mu}
\end{align}
\end{subequations}
where $V_{x}^{0}$ ($x=p,s$) is the natural wave velocity under strain-free state, and $dV_{x}/V_{x}^{0}$ represents the relative wave velocity change from strain-free state to a deformed state with thermal stain.

In the case of hydrostatic pressure, the strain caused by pressure $P$ is denoted as $\varepsilon_{P}$. For an isotropic body subject to uniaxial stress in direction 1, the principal strains are denoted as $\varepsilon_{1}$, and $\varepsilon_{2}=\varepsilon_{2}=-\nu\varepsilon_{1}$, where $\nu$ is the Poisson's ratio. 
The acoustoelastic coefficients in natural coordinates can also be derived from Eqs. (34)-(38) in Pao and Gamer's paper \cite{pao1985acoustoelastic}. For hydrostatic pressure,
\begin{subequations}
\label{eq:AE_hydro}
\begin{align}
     \frac{dV_p/V^{0}_p}{d\epsilon_{P}}&=
     1+\frac{3l+2m}{\lambda+2\mu}+ \frac{3\lambda+2\mu}{2\lambda+4\mu} \\
    \frac{dV_s/V^{0}_s}{d\epsilon_{P}}&=1+\frac{6m-n}{4\mu}+\frac{3\lambda+2\mu}{2\mu}
\end{align}
\end{subequations}
and for uniaxial stress,
\begin{subequations}
\label{eq:dv uniaxial l m n}
\begin{align}
    \frac{dV_{11}/V^{0}_{11}}{d\epsilon_{1}}&=\frac{3}{2}+\frac{l+2m-(\lambda+2l)\nu}{\lambda+2\mu}\\
    \frac{dV_{12}/V^{0}_{12}}{d\epsilon_{1}}&=1+\frac{\lambda+m-(2\lambda+2\mu+2m-n/2)\nu}{2\mu}\\
    \frac{dV_{21}/V^{0}_{21}}{d\epsilon_{1}}&= \frac{dV_{12}/V^{0}_{12}}{d\epsilon_{1}}\\
    \frac{dV_{22}/V^{0}_{22}}{d\epsilon_{1}}&= -\frac{3\nu}{2}+ \frac{\lambda+2l-(\lambda+4l+4m)\nu}{2\lambda+4\mu}\\
    \frac{dV_{23}/V^{0}_{23}}{d\epsilon_{1}}&=-2\nu+\frac{\lambda+m-n/2-2(\lambda+m)\nu}{2\mu}
\end{align}
\end{subequations}
where $V_{ij}^{0}$ is the natural wave velocity under stress-free state, $dV_{ij}/V^{0}_{ij}$ represents the relative wave velocity change from stress-free state to uniaxial stress state. The subscript $i$ stands for the wave propagation direction and $j$ for the polarization direction.

It should be noted that Eqs. \eqref{eq:AE_hydro} and \eqref{eq:dv uniaxial l m n} derived under natural coordinates have different forms from the initial (actual) velocity equations presented in many papers \cite{hughes1953second,egle1976measurement}. Conversion between natural and initial wave velocities can be found in Pao and Gamer's paper \cite{pao1985acoustoelastic}, and also discussed in detail by Duquennoy et al. \cite{duquennoy_influence_1999}  Eq. \eqref{eq:dv uniaxial l m n} also shows that the shear wave velocities $V_{12}$ and $V_{21}$ in natural coordinates show symmetry in polarization and propagation directions\cite{pao1985acoustoelastic}. 

Compared to Eq. \eqref{eq:dVp and dVs thermal expansion using l m n},  Eq. \eqref{eq:AE_hydro} for the hydrostatic pressure case has a third term that corresponds to the stress effect. Since TOEC have negative values for most materials, while $\lambda$ and $\mu$ are positive, both P and S wave velocities are more sensitive to temperature effect (Eq. \eqref{eq:dVp and dVs thermal expansion using l m n}) than to hydrostatic pressure.  Similar conclusions can also be drawn for comparison between temperature change and uniaxial loading (Eq. \eqref{eq:dVp and dVs thermal expansion using l m n} vs.  Eq. \eqref{eq:dv uniaxial l m n}). 

In Eqs. \eqref{eq:dVp and dVs thermal expansion using l m n}-\eqref{eq:dv uniaxial l m n}, Lam\'e constants $\lambda$ and $\mu$ can be determined from the wave velocities $V_p$, $V_s$ and density $\rho_0$ under strain-free state. In the case of uniaxial stress, the TOEC $l$, $m$ and $n$ can then be expressed by three acoustoelastic coefficients and Lam\'e constants. 
For uniform temperature change in isotropic material,\replaced{}{ similar to the case of hydrostatic pressure \cite{hughes1953second},} only two linear combinations of the three TOEC can be determined from two equations in Eq. \eqref{eq:dVp and dVs thermal expansion using l m n}, which are  
$C_{1}=l+n/9$ and $C_{2}=3l/5+m-n/10$, where $C_{1}$ is dominated by $l$, and $C_{2}$ is dominated by $m$.
\replaced{These two combinations of TOEC are two of three invariants of the TOEC defined in Blaschke's study \cite{blaschke2017averaging}:  $C_{1}=\frac{1}{54} C_{iijjkk},\, C_{2}=\frac{1}{30}C_{ijijkk}$,
where $C_{ijklmn}$ is the tensor of  the third-order elastic moduli \cite{pao1985acoustoelastic}.
}{These two combinations of TOEC were also defined by Blaschke \cite{blaschke2017averaging}.}
In thermal modulation tests, $C_{1}$ and $C_{2}$ can be derived from Eq. \eqref{eq:dVp and dVs thermal expansion using l m n} as:
\begin{subequations}
\label{eq:C1 and C2 thermal}
\begin{align}
    C_{1}&=\frac{\lambda+2\mu}{3}\left(\frac{dV_p/V^{0}_p}{d\epsilon_{T}}-1\right)-\frac{4\mu}{9}\left(\frac{dV_s/V^{0}_s}{d\epsilon_{T}}-1\right)\\
    C_{2}&=\frac{\lambda+2\mu}{5}\left(\frac{dV_p/V^{0}_p}{d\epsilon_{T}}-1\right)+\frac{2\mu}{5}\left(\frac{dV_s/V^{0}_s}{d\epsilon_{T}}-1\right)
\end{align}
\end{subequations}

To validate the proposed TOEC measurement method, two experiments were designed in this study: uniaxial loading test and the thermal modulation test. An aluminum 6061 block with a square cross section (5 cm $\times$ 5 cm) and a height of 15 cm was used in both tests. The sample has a density 2705 kg/$\text{m}^3$, and the Poisson's ratio 0.33. The velocities of P wave and S wave were measured as 6425 m/s and 3104 m/s under stress-stree condition. Then $\lambda$ and $\mu$ are calculated as $\lambda$ = 59.5 GPa and $\mu$= 26.1 GPa. The thermal expansion coefficient for aluminum is $\alpha_{T}=23$ $\mu\epsilon/^\circ\text{C}$.
 
\begin{figure}[h]
    \centering
    \includegraphics[width=6.5cm]{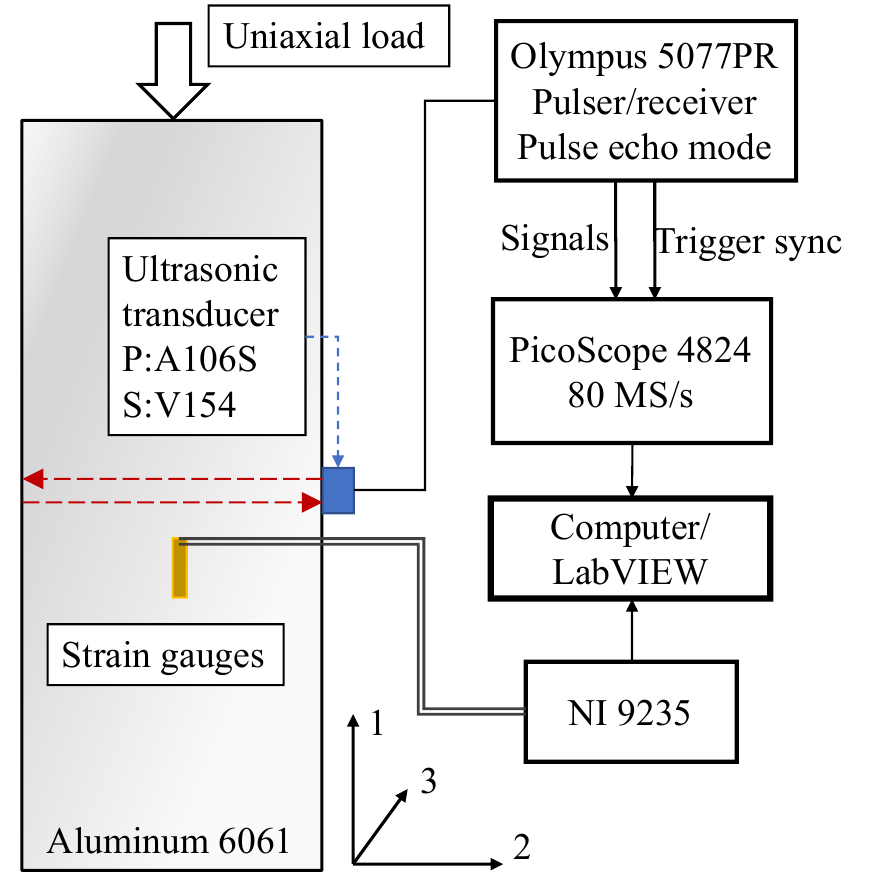}
    \caption{Experimental setup of uniaxial loading test.}
    \label{fig:setup uniaxial loading}
\end{figure}

Figure \ref{fig:setup uniaxial loading} shows the experimental setup for the uniaxial stress test, where the load was applied in the height direction. Two strain gauges were attached on both sides of the aluminum block to measure the axial strain in direction 1. A hydraulic compression machine was used to apply compression load until the strain reached about 250 $\mu\epsilon$. An ultrasonic transducer was driven by a pulser/receiver (Olympus 5077PR) in the pulse-echo mode to measure echo waves in direction 2. The loading test repeated three times, and in each loading test, one of the following ultrasonic transducers was used: (1) 2.25 MHz P wave transducer (Olympus A106S), (2) 2.25 MHz shear wave transducer (Olympus V154) polarized in direction 1, and (3) V154 polarized in direction 3. \added{The entire experimental process was completed at a room temperature of 23 $^\circ\text{C}$.} 
 
For common metal materials (aluminum, steel), the relative velocity changes are on the order of $10^{-6}/\mu\epsilon$. 
To accurately measure such small velocity changes, coda wave interferometry (CWI) \cite{lobkis2003coda} was used to calculate the dilation (stretching factor) between two signals. The stretching factor that maximizes the correlation coefficient represents the relative velocity change between two signals. With a sampling rate of 80 MS/s and using the CWI analysis method, we could measure $dV/V^{0}$ with a resolution of $10^{-6}$.

\begin{figure}[h]
    \centering
    \includegraphics[width=8cm]{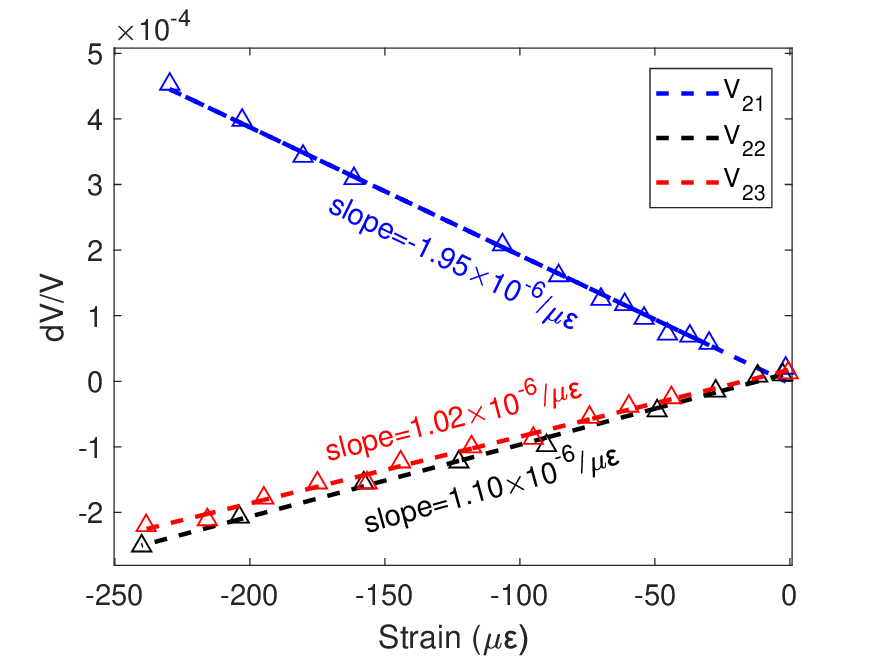}
    \caption{Velocity change with uniaxial strain for an aluminum 6061 sample.}
    \label{fig:dv/v uniaxial loading}
\end{figure}

Figure \ref{fig:dv/v uniaxial loading} presents the results for wave velocity changes under uniaxial stress/strain in direction 1. It can be seen that shear wave velocity $dV_{21}/V^{0}_{21}$ increases with the compression strain when its polarization aligns with the stress, while $dV_{22}/V^{0}_{22}$ and $dV_{23}/V^{0}_{23}$ decrease with the strain. The slopes give the acoustoelastic coefficients for each type of wave under uniaxial stress.

Thermal modulation tests were conducted on the same aluminum sample. Two ultrasonic transducers, V106S and V154, were installed at one end of the height direction (15 cm) to transmit and measure P and S waves. A switch (Agilent 34970A/34901A) was added before the pulser/receiver to switch between two transducers during the thermal process.  The temperature was monitored using a type T thermocouple and a data logger TC-08. In this test, the sample was first heated from 22 $^\circ\text{C}$ to 35 $^\circ\text{C}$, and then cooled down to 22 $^\circ\text{C}$ at a temperature changing rate of 1 $^\circ\text{C}/\text{hour}$. This rate of temperature change can avoid temperature gradient and ensure uniform thermal expansion in the sample. The temperature range gave about 300 $\mu\varepsilon$ thermal strain in the aluminum sample.

\begin{figure}[h]
    \centering
    \includegraphics[width=8cm]{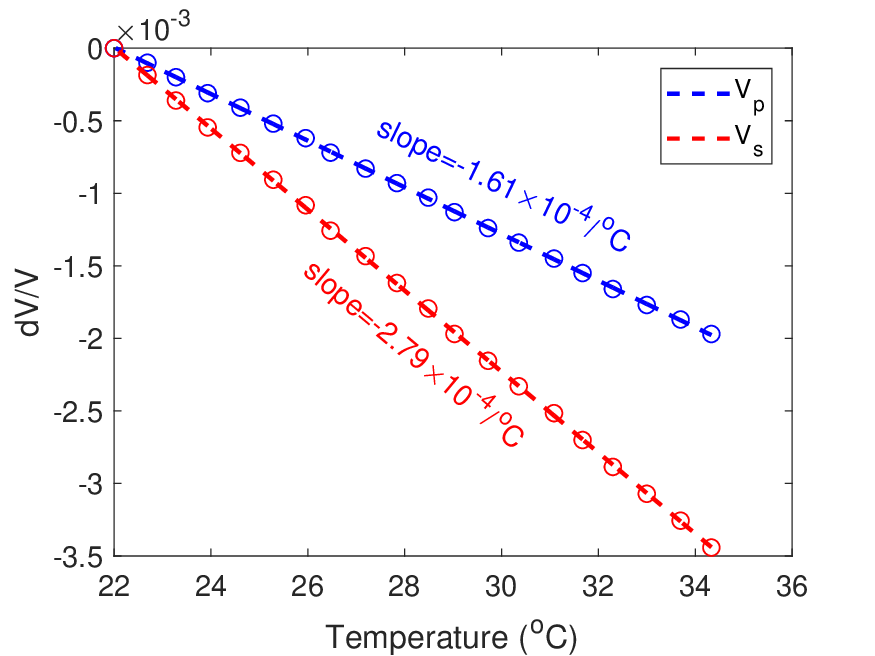}
    \caption{Thermal modulation test results on aluminum 6061 sample.}
    \label{fig:dv/v cooling}
\end{figure}

The data during the cooling process is used for analysis to match the negative strain in the uniaxial loading test. Figure \ref{fig:dv/v cooling} presents the results for relative P wave and S wave velocity changes due to temperature change. The data show high linearity with temperature and less variation than in the uniaxial test. The result indicates that the S wave velocity is more sensitive to temperature change than the P wave velocity, consistent with Weaver and Lobkis' results \cite{weaver2000temperature}.

\begin{table}[h]
    \centering
    \caption{Summary of acoustoelastic coefficients and TOEC}
    \label{tab:table1}
    \begin{adjustbox}{width=8.5cm}
    \begin{threeparttable}
    \begin{tabular}{@{}cccccccc@{}}
    \toprule
    \multicolumn{3}{c}{Uniaxial loading   } &  & \multicolumn{2}{c}{Thermal modulation}\\
    \midrule
    \large$\frac{dV_{21}/V^{0}_{21}}{d\epsilon_{1}}$ & \large$\frac{dV_{22}/V^{0}_{22}}{d\epsilon_{1}}$ & \large$\frac{dV_{23}/V^{0}_{23}}{d\epsilon_{1}}$ &  & \large$\frac{dV_p/V^{0}_p}{d\epsilon_{T}}$ & \large$\frac{dV_s/V^{0}_s}{d\epsilon_{T}}$  
    \\ \cmidrule(lr){4-4}
    -1.95 & 1.10 & 1.02 &  & -7.02 & -12.13&\\ \midrule
    $l$ & $m$  & $n$ &  & $C_{1}$ & $C_{2}$ \\ \cmidrule(lr){4-4}
    -113 & -298 & -337 &  & -146(-150)\tnote{*} & -316(-332)\tnote{*} \\ \bottomrule
    \end{tabular}%
    \begin{tablenotes}\footnotesize
    \item[*] Values of $C_{1}$, $C_{2}$ in parenthesis were calculated using $l$, $m$ and $n$ from the uniaxial loading test. All TOEC have the unit of GPa.
    \end{tablenotes}
    \end{threeparttable}
    \end{adjustbox}
\end{table}

Table \ref{tab:table1} summarizes the acoustoelastic coefficient from the uniaxial loading test and the thermal modulation test. The TOEC $l$, $m$ and $n$ can be calculated from Eq. \eqref{eq:dv uniaxial l m n} using the uniaxial test results, and $C_{1}$ and $C_{2}$ are calculated from  Eq. \eqref{eq:C1 and C2 thermal} using the thermal modulation test results and $\alpha_{T}=23$ $\mu\epsilon/^\circ\text{C}$. For comparison, $C_{1}$ and $C_{2}$ are also calculated based on $l$, $m$ and $n$ obtained from the uniaxial test. 
As shown in Table \ref{tab:table1}, these two independent tests give very close values on $C_{1}$ and $C_{2}$, with difference less than 5\%. This difference is less than the experimental errors reported in literature \cite{hughes1953second,blaschke2017averaging}, which validates the proposed thermal modulation method for TOEC measurements. 

As predicted from the equations, the relative velocity changes due to the thermal effect are larger than those caused by uniaxial loading. With the obtained $l$, $m$ and $n$, the P wave velocity change in uniaxial test can be calculated as ${dV_{11}/V^{0}_{11}}/{d\epsilon_{1}} = -4.35$, which is still less than the P wave acoustoelastic coefficient -7.02 in the thermal modulation test. This phenomenon was observed in a prior study of thermal modulation test \cite{sun2020determination}, where the authors noticed that the nonlinear parameters determined from the thermally induced velocity changes were larger than literature reported values, but no clear explanation was given. This letter provides a solid theoretical explanation of the high sensitivity of wave velocities to temperature change. 

The acoustic nonlinearity parameter $\beta$ is usually obtained from the second harmonic generation (SHG) test, by measuring the strain amplitudes at the fundamental and second harmonic frequencies after the wave travels a certain distance. In theory, it is directly related to the acoustoelastic coefficient of P wave in one-dimensional solids, $2{dV_p/V^{0}_p}/{d\epsilon}$.  It is challenging to accurately measure the absolute value of $\beta$, using either the SHG or uniaxial stress test. The parameter \replaced{$\beta=-3-(2l+4m)/(\lambda+2\mu)$}{$\beta=3+(2l+4m)/(\lambda+2\mu)$} actually represents the contribution of TOEC to the relative P wave velocity change. Similarly, we may define two new nonlinearity parameters based on Eq. \eqref{eq:C1 and C2 thermal}, which represent the contribution TOEC to the relative velocity changes of P and S waves. These two parameters are easy to measure using the thermal modulation test.


Compared to the uniaxial loading test, the thermal modulation test has a very simple experimental setup and produces uniform thermal strains. Because the test sample remains isotropic during the test, the polarization direction of the shear wave does not influence the velocity, and no polarization calibration is needed. In stress-free conditions, only two TOEC combinations $C_{1}$ and $C_{2}$ can be obtained in the thermal modulation test. In theory, it is feasible to determine all TOEC $l$, $m$ and $n$ by adding one more measurement through restricting expansion in one direction.

This letter presents the equations and experimental procedures to obtain acoustoelastic coefficients and TOEC of isotropic materials using thermally induced wave velocity changes. Analysis shows that the acoustoelastic coefficients $dV/V/d\varepsilon$ are more sensitive to the thermal effect than the stress effect. This conclusion is verified by experimental data obtained in the uniaxial test and the thermal modulation test.  Both tests give very close TOEC parameters. The thermal modulation test only needs a conventional ultrasonic measuring system and a temperature change environment. Compared to other nonlinear experimental methods, the thermal modulation test may have improved reliability and robustness due to its simple test setup and the large values of thermo-acoustoelastic coefficients. \added{Prior studies by Sun and Zhu \cite{sun2019thermal,sun2020determination} present the feasibility of using the thermal modulation test to characterize concrete with microcracking damage. Further studies are needed to investigate the sensitivity of thermally induced nonlinearity to damage in materials. }

\begin{acknowledgments}
This research was partially supported by the U.S. Department of
Energy-Nuclear Energy University Program (NEUP) under
Contract No. DE-NE0008544.
\end{acknowledgments}

\section {DATA AVAILABILITY}
The data that support the findings of this study are available
from the corresponding author upon reasonable request.

\bibliographystyle{aipnum4-1} 
\bibliographystyle{apsrev4-1}
\bibliography{aipsamp}

\end{document}